\documentclass[12pt]{article}

\newcommand{\eps}{\varepsilon}
\newcommand{\ep}{\varepsilon^\prime}
\newcommand{\epp}{\varepsilon^\prime/\varepsilon}
\newcommand{\epsp}{\frac{\varepsilon^\prime}{\varepsilon}}
\newcommand{\be}{\begin{equation}}
\newcommand{\ee}{\end{equation}}
\newcommand{\bea}{\begin{eqnarray}}
\newcommand{\eea}{\end{eqnarray}}
\newcommand{\nn}{\nonumber}
\newcommand{\DSone}{\Delta S=1}

\begin{document}
\pagestyle{empty}
\setcounter{page}{1}
\begin{center}
{\normalsize\sf
\rightline{TUM-HEP-376/00}
\rightline{RM3-TH/00-10}
}
\vskip 2cm
{\Large\bf Theoretical status of \boldmath$\epp$$\,^\star$}
\vskip 1cm
{\large Marco Ciuchini$\,^a$ and Guido Martinelli$\,^b$}\\
\vskip 0.5cm
\baselineskip=14.5pt
$^a${\small\em Physik Dept., Technische Universit{\"a}t M{\"u}nchen,
D-85748 Garching, Germany.}\\
{\small\em Dip. di Fisica, Universit{\`a} di Roma Tre
 and INFN, Sezione di Roma III,}\\
{\small\em Via della Vasca Navale 84, I-00146 Rome, Italy.}\\
\vskip 0.3cm
$^b${\small\em Laboratoire de Physique Th\'eorique (LPT),
 Universit\'e de Paris-Sud,}\\
{\small\em B\^atiment 210, 91405 Orsay.} \\
{\small\em Centre de Physique Th\'eorique de l'\'Ecole Polytechnique,}\\
{\small\em 91128 Palaiseau Cedex, France.}
\end{center}
\vskip 1cm
\begin{abstract}
We review the theory of $\epp$ and present an updated phenomenological
analysis using hadronic matrix elements from lattice QCD.
The present status of the computation of $\epp$, considering various
approaches to the matrix-element evaluation, is critically discussed.
\end{abstract}
\vfill
\rule{5cm}{0.5pt}\\
{\footnotesize $^\star$ based on the talks
given by M.C. at ``Les Rencontres de Physique de la Vall\'ee d'Aoste'',
La Thuile (Italy), 27 February--4 March 2000 and by G.M. at the
``XXXVth Rencontres de Moriond'', Les Arcs 1800 (France), 11--18 March 2000.}

\baselineskip=17pt
\newpage

\pagestyle{plain}
\setcounter{page}{1}

\section{Introduction}
The latest-generation experiments, aiming to obtain $\epp$
with a $10^{-4}$ accuracy, measured up to now
\be
\epsp=\Bigg\{
\begin{tabular}{l l}
$(28.0\pm 4.1)\times 10^{-4}$ & KTeV~\cite{KTeV}\\
$(14.0\pm 4.3)\times 10^{-4}$ & NA48~\cite{NA48}
\label{eq:expres}
\end{tabular}
\ee
By combining these results with previous measurements,
the latest world average reads~\cite{NA48}
\be
\epsp=(19.3\pm 2.4)\times 10^{-4}\,,
\ee
which is definitely in the $10^{-3}$ range. 
Given the differences in the results of eq.~(\ref{eq:expres}), 
the quoted error is, however,  debatable~\cite{dagostini}.

On the other hand, theoretical estimates in the Standard-Model typically
correspond to  central values  in
the $10^{-4}$ range although, given the large theoretical uncertainties,
values of the order of  $10^{-3}$ are  not excluded. 
The explanation of the difference  between SM
predictions and experimental values calls either for 
some missing
dynamical effect in the hadronic parameters or for physics beyond the Standard Model.
In the last few months, several studies exploring both possibilities have been
published.

In this paper,  the theoretical status  of $\epp$ is reviewed 
and updated results obtained by  using (whenever possible) hadronic 
matrix elements
computed with lattice QCD are presented.  Other theoretical approaches and
recent attempts to ``improve'' the accuracy in the determination
of the hadronic matrix elements (mostly to improve the 
agreement between theoretical estimates and measurements) are also discussed.

\section{Basic formulae}

Direct CP violation, occurring in $K^0$ decays, is parametrized by
$\ep$. In terms of  weak-Hamiltonian matrix
elements, this quantity is  defined as
\be
\ep=
\frac{\langle \pi\pi(0)\vert H_W\vert K_S\rangle\langle\pi\pi (2)\vert
H_W\vert K_L\rangle-\langle \pi\pi(0)\vert H_W\vert K_L\rangle
\langle \pi\pi(2)\vert H_W\vert K_S\rangle}
{\sqrt{2}\langle \pi\pi(0)\vert H_W\vert K_S\rangle^2}\,,
\ee
where the $\langle\pi\pi(I)\vert$ is the isospin $I$ two-pion {\it out}-state and
\bea
\vert K_L\rangle &=& \frac{\vert K^0\rangle-\vert \bar K^0\rangle +
\bar\eps (\vert K^0\rangle+\vert \bar K^0\rangle) }
{\sqrt{2(1+\vert \bar\eps\vert^2)}}\,\nn\\
\vert K_S\rangle &=& \frac{\vert K^0\rangle+\vert \bar K^0\rangle  +
\bar\eps (\vert K^0\rangle-\vert \bar K^0\rangle) }
{\sqrt{2(1+\vert \bar\eps\vert^2)}}
\eea
are the eigenstates of the  CPT-conserving Hamiltonian describing the
$K^0$--$\bar K^0$ system, namely
\be
H_{K^0-\bar K^0}=M-\frac{i}{2}\Gamma=\left(\begin{array}{cc}
M_0 & M_{12}\\M_{12}^* & M_0
\end{array}\right)
-\frac{i}{2}\left(\begin{array}{cc} \Gamma_0 & \Gamma_{12}\\
\Gamma_{12}^* &\Gamma_{0}\end{array}\right)\, .
\ee

We introduce the isospin amplitudes
\be
A_Ie^{i\delta_I}=\langle \pi\pi (I)\vert H_W\vert K^0\rangle\, ,\qquad
A_I^\star e^{i\delta_I}=\langle \pi\pi (I)\vert H_W\vert \bar K^0\rangle\, ,
\ee
where, in virtue of Watson's theorem, 
the  $\delta_I$s are the strong-interactions phase shifts
of $\pi\pi$ scattering.
In  the approximation $\mbox{Im}A_0\ll \mbox{Re}A_0$, $\mbox{Im}A_2\ll \mbox{Re}A_2$
and $\omega=\mbox{Re}A_2/\mbox{Re}A_0\ll 1$ (the latter coming from the
$\Delta I=1/2$ enhancement in kaon decays), one finds 
\bea
\eps &\simeq& \bar\eps+i\frac{\mbox{Im}A_0}{\mbox{Re}A_0}\simeq\frac{e^{i\pi/4}}{\sqrt{2}}\left(\frac{\mbox{Im}M_{12}}
{2\mbox{Re}M_{12}}+\frac{\mbox{Im}A_0}{\mbox{Re}A_0}\right)\, ,\nn\\
\ep &\simeq& i\frac{e^{i(\delta_2-\delta_0)}}{\sqrt{2}}\mbox{Im}\left(
\frac{A_2}{A_0}\right)
\simeq i\frac{e^{i(\delta_2-\delta_0)}}{\sqrt{2}}\frac{\omega}{\mbox{Re}A_0}
\left(\omega^{-1}\mbox{Im}A_2-\mbox{Im}A_0\right)\, .
\eea
Using the experimental value~\cite{d0d2}
\be
\mbox{Arg}\,\ep=\frac{\pi}{2}+\delta_2-\delta_0=(48\pm 4)^\circ\approx \pi/4\, ,
\ee
one finally gets
\be
\epsp\simeq\frac{1}{\sqrt{2}\vert\eps\vert}\frac{\omega}{\mbox{Re}A_0}
\left(\omega^{-1}\mbox{Im}A_2^\prime-(1-\Omega_{IB})\mbox{Im}A_0\right)\, ,
\label{eq:epsp0}
\ee
where the last expression includes   isospin breaking contributions 
due to $\pi$--$\eta$ mixing encoded in  $\Omega_{IB}$ 
($A_2^\prime = A_2-\omega\Omega_{IB} A_0$)~\cite{bg87}. In the prediction of
$\epp$, $\omega$ and $\mbox{Re}A_0$ are taken from  experiments, whereas
 $\mbox{Im}A_{0,2}$ are the computed quantities.

The calculation of the real part of  the amplitudes, and hence of $\omega$,
is one of the longest-standing problems in particle physics:
in spite of several decades of efforts,
nobody succeeded so far  to  explain the $\Delta I=1/2$ rule
in a convincing and  quantitative way.  The calculation of $\mbox{Im}A_0$ 
and $\mbox{Im}A_2$  is of comparable difficulty. Since 
the imaginary parts  entering $\epp$, however, are not directly related
to the real ones, and the operators of the effective Hamiltonian
contribute  with different weights in the two cases,
it is conceivable that $\mbox{Im}A_0$  and $\mbox{Im}A_2$ be computed in spite
of the difficulties encountered in calculations of the $\Delta I=1/2$ rule. On the
other hand, as 
discussed below, one cannot exclude some  common dynamical enhancement
mechanism which produces large values of both $\mbox{Re}A_0$ and  $\epp$.

\section{\boldmath$\epp$  in the Standard Model}
The natural theoretical framework in dealing with  weak hadronic decays
is provided by the effective Hamiltonian formalism. Indeed the operator product
expansion allows the separation of short- and long-distance scales and
reduce the problem to the computation of Wilson coefficients, performed in
perturbation theory, and to the calculation, with non-perturbative techniques,
of {\em local}-operator matrix elements. 

 At the next-to-leading order (NLO) in the renormalization-group
improved expansion, the 4-active-flavour ($m_b > \mu > m_c$) $\DSone$ effective
Hamiltonian, relevant for  $\epp$, can be written as
\bea
H_W &=&-\frac {\lambda_u G_F} {\sqrt{2}}
\Bigl\{ (1 - \tau ) \Bigl[ C_1(\mu)\Bigl( Q_1(\mu)- Q_1^c(\mu) \Bigr)
+ C_2(\mu)\left( Q_2(\mu) - Q_2^c(\mu) \right)  \Bigr]\nn\\
&~&+ \tau \sum_{i=1}^{9} C_i(\mu) Q_i(\mu) \Bigr\}\,~,
\label{eq:heff}
\eea
where $G_F$ is the Fermi constant, $\lambda_q=V_{qd} V^\star_{qs}$ and
$\tau=-\lambda_t/\lambda_u$ ($V_{q_i q_j}$ being the CKM matrix elements).
The $CP$-conserving and $CP$-violating contributions are easily separated,
the latter being proportional to $\tau$.

Neglecting electro- and chromo-magnetic dipole transitions, the operator basis
includes eleven independent local four-fermion operators. They are given by
\be
\begin{array}{ll}
Q_{ 1} =({\bar s}_{\alpha}d_{\alpha})_{V-A}
    ({\bar u}_{\beta}u_{\beta})_{V-A}\, ,
   &
Q_{ 2} =({\bar s}_{\alpha}d_{\beta})_{V-A}
    ({\bar u}_{\beta}u_{\alpha})_{V-A}\, ,
   \\
Q_{ 1}^c =({\bar s}_{\alpha}d_{\alpha})_{V-A}
    ({\bar c}_{\beta}c_{\beta})_{V-A}\, ,
   &
Q_{ 2}^c =({\bar s}_{\alpha}d_{\beta})_{V-A}
    ({\bar c}_{\beta}c_{\alpha})_{V-A}\, ,
   \\
Q_{3,5} =
    ({\bar s}_{\alpha}d_{\alpha})_{V-A}
    \sum_q({\bar q}_{\beta}q_{\beta})_{V\mp A}\, ,
   &
Q_{4,6} = ({\bar s}_{\alpha}d_{\beta})_{V-A}
    \sum_q({\bar q}_{\beta}q_{\alpha})_{V\mp A}\, ,
   \\
Q_{7,9} = \frac{3}{2}({\bar s}_{\alpha}d_{\alpha})_
    {V-A}\sum_{q}e_{ q}({\bar q}_{\beta}q_{\beta})_
    {V\pm A}\, ,
   &
Q_{8} = \frac{3}{2}({\bar s}_{\alpha}d_{\beta})_
    {V-A}\sum_{q}e_{ q}({\bar q}_{\beta}q_{\alpha})_
    {V+A}\, ,\\
\end{array}
\label{eq:opbasis}
\ee
where $(\bar q_\alpha q^\prime_\beta)_{V\pm A}=\bar q_\alpha\gamma_\mu
(1\pm\gamma_5)q^\prime_\beta$, $\alpha$ and $\beta$ are colour indices,
and the sum index $q$ runs over $\{d,u,s,c\}$.
The operator $Q_2$ appears in the Fermi Hamiltonian at tree level.
The operators $Q_3$--$Q_6$ are generated by the insertion of 
 $Q_2$ into the strong penguin diagrams, whereas $Q_7$--$Q_9$ come
from the electromagnetic penguin diagrams. Both
classes of operators are  relevant for $\epp$.
Further details on the NLO $\DSone$ effective Hamiltonian can be found
in refs.~\cite{ds1}.

Using eq.~(\ref{eq:heff}), one can readily express $A_0$ and $A_2$
in terms of matrix elements of the operators in eq.~(\ref{eq:opbasis}).
It is customary to write
\bea
\mbox{Im}A_0 &=&-\frac{G_F}{\sqrt{2}} \mbox{Im}\Bigl(V_{ts}^\star V_{td}
\Bigr)\Bigl\{-\Bigl(C_6 B_6+\frac{1}{3}C_5 B_5\Bigr)Z
+\Bigl(C_4 B_4+\frac{1}{3}C_3 B_3 \Bigr)X\nn\\
&&+C_7 B_7^{1/2}\Bigl(\frac{2Y}{3}+\frac{Z}{6}+\frac{X}{2}\Bigr)
+C_8 B_8^{1/2}\Bigl(2Y+\frac{Z}{2}+\frac{X}{6}\Bigr)\nn\\
&&-C_9 B_9^{ 1/2}\frac{X}{3}
+\Bigl(\frac{C_1 B_1^c}{3}+C_2 B_2^c\Bigr)X\Bigr\}~,\nn\\
\\
\mbox{Im}A_2^\prime &=& -G_F\mbox{Im}\Bigl(V_{ts}^\star V_{td}\Bigr)
\Bigl\{C_7 B_7^{3/2}\Bigl(\frac{Y}{3}
-\frac{X}{2}\Bigr)+C_8 B_8^{3/2}\Bigl(Y-\frac{X}{6}\Bigr)\nn\\
& & +C_9 B_9^{3/2}\frac{2X}{3}\Bigr\}\, .\nn
\label{eq:ima0a2}
\eea
In the previous equation, the relevant matrix elements are given in
terms of $B$-parameters  defined as
\bea
\langle \pi\pi(0)\vert Q_i\vert K\rangle &=&
B^{1/2}_i\langle \pi\pi(0)\vert Q_i\vert K\rangle _{VIA}\, ,\nn\\
\langle \pi\pi(2)\vert Q_i\vert K\rangle &=&
B^{3/2}_i\langle \pi\pi(2)\vert Q_i\vert K\rangle _{VIA}\, ,
\label{eq:bpars}
\eea
where the subscript $VIA$ means that the matrix elements are calculated in
the vacuum insertion approximation. $VIA$ matrix elements are given
in terms of the three quantities
\bea
X\!&=&\!f_{\pi}\left(M_{ K}^{ 2}-M_{\pi}^{ 2}\right)~,\nn\\
Y\!&=&\!f_{\pi}\left(\frac{M_{ K}^{ 2}}{m_s(\mu)+m_d(\mu)}\right)^2
 \sim 12\,X\left(\frac{150 \, \mbox{MeV}}{m_s(\mu)}\right)^2~,\\
Z\!&=&\!4\left(\frac{f_{ K}}{f_{\pi}}-1\right)Y~.\nn
\label{eq:xyz}
\eea
Contrary to $X$ and $Z$, $Y$ does not vanish in the chiral
limit, as a consequence of the different chiral properties of the operators
$Q_7$ and $Q_8$. Moreover, whereas $X$ is expressed in terms of measurable
quantities, both $Z$ and $Y$ depend on the  quark masses which must be taken from
theoretical estimates.

Note that some $VIA$ matrix elements seems to show a quadratic dependence on the
strange quark mass $m_s$ through $Y$ and $Z$. This is true 
 as long as one fixes the kaon mass to its experimental value and neglect the
$m_s$ dependence of the $B$ parameters. The apparent quadratic
dependence of the matrix elements on $m_s$  has been exploited
in ref.~\cite{kns99} to  claim that large values of $\epp$ can
be obtained with suitably ``small''  strange quark masses. 
The actual dependence of the full matrix elements on $m_s$ is, however, 
very different.
Indeed the ratio $M_K^2/m_s$ is essentially independent of $m_s$ (up to small
chiral-symmetry-breaking terms)   since it corresponds
to the value of quark condensate. 
This is explicitly verified  in   lattice calculations,
where  a strong correlation between the value of the strange quark
mass used in the $VIA$ matrix elements and the value of the
corresponding $B$ parameters is observed, so that the $m_s$ dependence
in the physical matrix element almost cancels out~\cite{giusti}.
 This is why one should always use
$B$-parameters and $m_s$ consistently computed together ({\it e.g.}
in the same simulation on the lattice) or, even better, matrix elements
given in physical units without any reference to quark masses~\cite{giusti,k99}.

\section{Hadronic matrix elements calculation}
Any  prediction  of $\epp$ must undergo
the non-perturbative calculation of the relevant hadronic matrix elements.
Theoretically, this calculations has  to meet two requirements
\begin{enumerate}
\item to be applicable up to perturbative energy scales;
\item to keep under control the definition of the renormalized operators and their
consistent  matching to the Wilson coefficients.
\end{enumerate}
Failure to meet these requirements indicates that the method cannot achieve
the necessary NLO accuracy, as often the case with phenomenological models.
Presently, however (or may be for this reason),
experimental data are more easily accommodated by such models than by methods based on
first principles, as lattice QCD. 

\begin{figure}[t]
\vspace{9.0cm}
\includegraphics{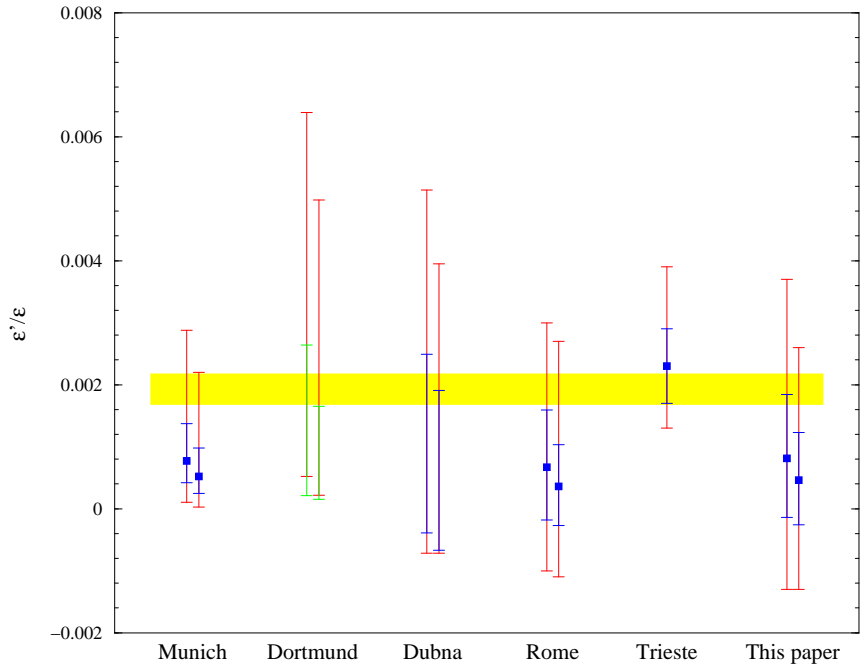}
 \caption{\it
      Compilation of recent theoretical predictions for $\epp$. 
    \label{fig:comp} }
\end{figure}

With this caveat in mind, we list and briefly comment on various approaches that
have been used in the literature.
Various  predictions of $\epp$, obtained using different methods, are shown
in the compilation of fig.~\ref{fig:comp}, together with the present experimental world average.

\subsubsection*{Lattice QCD~\cite{rom99}}
In principle, {\it Lattice QCD is the non-perturbative method for computing matrix
elements}. Being a regularized version of the fundamental theory, it allows
a complete control over the definition of the renormalized operators, both at the
perturbative and non-perturbative level. In addition, present simulations use
inverse lattice spacings of $2$ GeV or larger and therefore
the  perturbative matching with the
Wilson coefficients can be safely performed. Indeed, by using non-perturbative
renormalization techniques, the matching scale in lattice calculations
could  be pushed to values as large as $\sim 10$ GeV. Although, so far, these
methods have only been implemented  in the calculation of the strong coupling
constant~\cite{luscher}
and of the  quark masses~\cite{wittig},   they will certainly be
extended to the four-fermion operators of the weak effective Hamiltonian.
When  this will be the case, the error in the matching
procedure will become negligible.

For many years,  a general no-go theorem~\cite{mt} of Euclidean field theory
has prevented the  direct extraction, in numerical simulations,
of the physical matrix elements with more than one particle in the final state.
For this reason, 
present lattice determinations of $\langle\pi\pi\vert Q_i\vert K\rangle$
are obtained from $\langle\pi\vert Q_i\vert K\rangle$ using lowest-order
chiral relations ({\it i.e.} soft-pion theorems). This means that  final-state
interactions are not taken into account and that large
chiral corrections may be present~\cite{ppp}~\footnote{ Other typical lattice systematics,
such as the quenching, are expected to play a lesser  r\^ole in this context.}.
In addition, only some of the matrix elements needed for computing $\epp$ are
presently available. In
particular, the $Q_6$ matrix element, which is expected to give the most important
contribution to $\epp$, has not been successfully computed yet.

Several theoretical progresses have opened a window of opportunity this  year.
In the past,  a proposal to circumvent the no-go theorem of 
ref.~\cite{mt} was made.  The main idea was to extract the relevant matrix elements
by studying suitable Euclidean Green functions at small 
time distances~\cite{rom95}. The weakness of this method, however,
 was that it relied  on  model-dependent  smoothness assumptions which could lead
 to uncontrolled systematic errors.     
A big progress has been made in ref.~\cite{ll},  where 
it was rigorously proven how to relate
the matrix elements  extracted on a
finite volume in lattice simulations to the
physical  $\langle\pi\pi\vert Q_i\vert K\rangle$  amplitudes.
Moreover,  it has been shown that 
the smoothness hypothesis of ref.~\cite{rom95} is unnecessary and 
that the  physical  $K \to
\pi\pi$ matrix elements can be, at least in principle,
extracted from Euclidean correlation functions
at finite-time distances~\cite{sacring}.
Although it will take some time before these approaches will be implemented in practice,
they certainly open new perspectives to lattice calculations.  

More details on the present status of lattice matrix elements, and possible developments
in the near future,  will be given
in the discussion of  the phenomenological analysis in the next section.

\subsubsection*{Phenomenological Approach~\cite{mun99}}
In this approach, one basically attempts to extract information on the matrix
elements relevant for $\epp$ by combining the measured values of the $CP$-conserving
amplitudes with relations among different operators that can be established below
the charm threshold under very mild assumptions (for details see ref.~\cite{mun93}).
This procedure can be performed consistently at the NLO, allowing 
the extraction of
matrix elements of well-defined renormalized operators.

Unfortunately, the leading contributions to $\epp$, namely the matrix elements
of $Q_6$ and $Q_8^{3/2}$,  cannot be  fixed  in this approach. Moreover,
the method only works below the charm threshold where higher-order perturbative and power
corrections (in $1/m_c$)  may be large. In practice, for these matrix elements, 
Buras and collaborators have always used inputs coming from other  theoretical sources,
in particular   lowest-order $1/N$ expansion or lattice calculations.

\subsubsection*{Chiral+$1/N$ Expansion~\cite{dor99}}
This method relies on the non-perturbative technique originally proposed by Bardeen,
Buras and G\'erard~\cite{bbg}. In principle, the approach can be derived from QCD and
allows  the computation of all
the matrix elements needed for calculating $\epp$ in a
consistent  theoretical scheme. In this framework, 
the Dortmund group computed the relevant matrix
elements including the subleading  corrections  in both the chiral and the
$1/N$ expansion~\footnote{ Indeed,
some of the higher-order terms have been computed in the chiral
limit only.}. 

This approach suffers, however, from the presence of
quadratic divergences in the cutoff that must be introduced,
beyond the leading order, in the effective
chiral Lagrangian.
The quadratic cutoff dependence, which  appears in non-factorizable contributions, 
makes it impossible 
a consistent matching between the operator matrix elements and the
corresponding Wilson coefficients,
which depend only logarithmically on the cutoff.
One may argue that  the quadratic divergences will be cured and replaced by 
some hadronic scale  in the full theory, 
which includes excitations heavier than the
pseudoscalar mesons.  
In practice, since it is impossible to include the effects
of higher-mass hadronic states, the cutoff is replaced with a  scale of the order
of 1 GeV, which is an arbitrary, although reasonable, choice.
Since the divergent terms  gives very 
large  contributions  to the
matrix elements entering $\Delta I=1/2$ transitions and
$\epp$, this introduces an uncontrolled numerical   uncertainty in the
final results.

\subsubsection*{Chiral Quark Model~\cite{tri00}}
The $\chi$QM  can be derived in the framework of the extended
Nambu-Jona-Lasinio model of chiral symmetry breaking~\cite{tri00}.
It contains an effective interaction between the $u$, $d$, $s$ quarks and the pseudo-scalar
meson octet with three free parameters, two of which can be fixed using
$CP$-conserving amplitudes.
The Trieste group computed the ${\cal O}(p^4)$ corrections to the relevant operators
and found a correlation between the $CP$-conserving and $CP$-violating amplitudes
so that, once the parameters of the model are fixed to provide the required octet
enhancement, it is possible to predict $\epp$.
 The nice feature of this model is that, to some extent, it accounts
for higher-order chiral effects, which are not easily included, for instance,
in lattice calculations.
The disadvantage is that the model dependence of the results 
can hardly be evaluated or corrected.

Theoretically, this approach shares some of the problems of the $1/N$ expansion
 {\it mainly} those related to the presence of 
quadratic divergences. These do not  appear explicitly in the calculations
of the Trieste group simply because dimensional regularization is used. 
It remains true, however, that  the
scale and scheme dependence of the renormalized operators is not  
under control at  NLO.
In order to deal with this problem, 
a third parameter of the model is fixed by imposing a sort of
 numerical ``$\gamma_5$''-independence to the
physical amplitudes.
This  recipe, while suggesting that
some degree of ``effective'' renormalization-scheme independence can be achieved,
has unfortunately  no sound theoretical basis. 
Finally the correlation between the $\Delta I=1/2$
amplitude  and $\epp$ is subject to  potentially large uncertainties for the
following reason.  The parameters necessary to estimate
the matrix element of  $Q_6$  are fixed
by fitting the $\Delta I=1/2$ amplitude. For this quantity, 
the  contribution of $Q_6$ is rather marginal, whereas $Q_1$ and $Q_2$ dominate.
Thus any small uncertainty in the dominant terms,  due 
for example  to  unknown ${\cal O}(p^6)$
corrections (${\cal O}(p^4)$ corrections to the $\Delta I=1/2$ amplitude
are of ${\cal O}(100\%)$), may change drastically
the determination  of the matrix element of $Q_6$, which is the dominant term 
for $\epp$.

\subsubsection*{Extended Nambu-Jona-Lasinio Model~\cite{bel99}}
An extended Nambu-Jona-Lasino model has been used in ref.~\cite{bel99} to 
compute the $\Delta I =1/2$
$K\to\pi\pi$ matrix elements and $\epp$. The remarkable feature of this computation
is the high order in the momentum expansion reached by the Dubna group.
All matrix elements have been computed  to ${\cal O}(p^6)$ and a good stability
 of the results
has been found. In this respect, this approach is safer than the
$\chi$QM. However it shares with the
$\chi$QM all the other theoretical flaws mentioned above,
 and particularly the problem of
matching the short-distance calculations, since it is
 unclear which renormalized
operators the amplitudes computed with the Dubna superpropagator regularization
method correspond to.

\subsubsection*{Generalized Factorization~\cite{che99}}
Generalized factorization has been introduced in the framework of non-leptonic
$B$ decays in order to parametrize the hadronic matrix elements without
{\em a-priori} assumptions~\cite{genfac}.
The basic idea is to extract from experimental data 
as much information as possible on the non-factorizable parameters. 
 When needed, the number of independent parameters can
be reduced using flavour symmetries, dynamical assumptions, etc. In ref.~\cite{che99}
the  procedure has been applied to $K\to\pi\pi$ matrix elements. Unfortunately,
in this case, the
number of independent channels that  one can use  to fix the parameters
is small (essentially only the two $CP$-conserving amplitudes). For his  predictions,
the author of ref.~\cite{che99} was forced then to reduce the number of parameters by
several   ``simplifying'' assumptions, which are,   however, questionable. Many parameters
related to different operator matrix elements  and to different colour structures were arbitrarily
assumed to be equal. In such a way, a correlation between $CP$-conserving and $CP$-violating
amplitudes was  obtained, but the final results  crucially depends on the assumptions,
 which are
hardly justifiable theoretically, and cannot be tested phenomenologically in processes different
from  $\epp$.

\subsubsection*{\boldmath$\sigma$ Models~\cite{kns99,har99}}
A possible mechanism to enhance the $\Delta I=1/2$ amplitude is the exchange of a scalar
$I=0$ meson~\cite{gavela}. It also leads to an enhancement of $\epp$, as recently studied in the framework
of the linear~\cite{kns99} and non-linear~\cite{har99} $\sigma$ models. While unable to achieve
NLO accuracy, these models can produce the required correlation between the $\Delta I=1/2$
rule and $\epp$, at least for some choice of the free parameters, such as the $\sigma$ mass.
Also in this case, however, it is not easy to estimate the uncertainties and the model dependence
of the theoretical  predictions.

\subsubsection*{Other theoretical developments}
The marginal agreement between  the SM predictions and the measured value of
$\epp$ stimulated various attempts to ``improve'' the determination of the
operator  matrix elements, by 
including  effects that were  not considered  previously.
In particular, new studies have been devoted to  the calculation 
of  isospin-breaking and final-state interaction effects.

In most of the approaches  isospin-breaking
corrections  are not  included because they are beyond reach for
these methods. These effects can be evaluated, however,
{\it a posteriori} and included in the predictions. The leading effect in the chiral expansion is expected to come from
$\pi$--$\eta$--$\eta^\prime$ mixing,  which can be computed following
ref.~\cite{bg87}.
The resulting isospin-breaking effect is accounted for by the parameter $\Omega_{IB}$
which appears in eq.~(\ref{eq:epsp0}). Recently the calculation of $\Omega_{IB}$
has been updated by  including the effect of $\pi^0-\eta$ mixing at ${\cal O}(p^4)$~\cite{eck99}.
In addition, it has been pointed out in ref.~\cite{gv99} that new sources of isospin breaking appear,
 beyond the leading order, in the
chiral Lagrangian. These terms   may give large corrections to
$\Omega_{IB}$. Unfortunately the calculation of the  corrections is strongly
model dependent and can only been taken as a warning on the potential importance  of these effects.

The problem of including final-state interactions is particularly relevant for lattice
or lowest-order $1/N$ calculations, where rescattering effects are missing.
It has recently been suggested that these effects could be included by using the measured
$\pi\pi$ phase shifts and dispersive techniques~\cite{ppp}.  The resulting $A_0$ amplitude
would be enhanced by the inclusion of final state interactions, giving for $\epp$ a prediction
much closer to the experimental value.
These approach have been subject 
to several criticisms~\cite{mun00}. On the one hand the analytic structure of the
considered amplitudes is unclear and the corresponding dispersion relations questionable.
On the other, the computation of the dispersive correction factors, as derived in ref.~\cite{ppp},
is plagued by an irreducible ambiguity of the same order
of the dispersive factors themselves. This uncertainty
depends on  the choice of the initial conditions which, as shown in~\cite{mun00},
were arbitrarily chosen. For this reason, whereas    final-state interactions are likely
to give qualitatively  a certain enhancement, as argued   in ref.~\cite{ppp}, the quantitative estimate
of these effect is subject to very large uncertainties.  As discussed in~\cite{mun00},
lattice calculations could help in this respect  by fixing the initial conditions in a unambiguous way.

Some short, provocative comment is necessary at this point.  If one could implement  
in the same calculation all the {\it corrections} which
have been suggested to improve the accuracy in the determination of the matrix elements
(low strange-quark mass, isospin-breaking effects, final state interactions, etc.), one would
probably end up with a prediction of $\epp$ much larger than its experimental value.
It is also quite astonishing that  the effects which were not considered before,
or those which have been
revised in recent studies, all increase the theoretical prediction for this quantity and no one goes in the
opposite direction.
Finally, if  the $\Delta I=1/2$ rule and the large value of $\epp$ are a consequence 
of many effects which  are all necessary in a  conspiracy   to give a large enhancement,
it seems very unlikely that any of the existing
theoretical approaches (including the lattice one), will ever be able to take them into account
simultaneously at the necessary level of accuracy.

During the completion of this paper, several new calculations of $\epp$ appeared:
i) a new estimate of the $Q_6$ and $Q_8^{3/2}$
matrix elements using QCD sum rules has been presented~\cite{nar00}, 
with  results for $\epp$  close to the experimental  average; ii) within big uncertainties,
very large values  of $\epp$ have also been obtained using the $1/N$ and chiral expansions within
the context of the extended Nambu-Jona-Lasinio model.  In ref.~\cite{xbos},
the  proposal for  controlling the
scale and scheme dependence of renormalized operators using an intermediate   $X$-boson
has been implemented in the calculation of $\epp$.
We refer the reader to the original publications for more details.

\section{Results for \boldmath$\epp$ using Lattice QCD}
In order to compute $\epp$, besides hadronic matrix elements
one needs  the value  of the relevant combination
of CKM matrix elements, namely Im$V_{ts}^\star V_{td}$. This is constrained 
by using the experimental information on $\vert V_{cb}\vert$, $\vert V_{ub}\vert$,
$B_{d,s}$--$\bar B_{d,s}$ mixings and $\eps$ combined with lattice results.
Nowadays this has become a standard
way of determining the CKM-matrix parameters within the Standard Model, described
for instance in refs.~\cite{rom99,ckm}. It is worth noting that the linear
dependence of $\epp$ on Im$V_{ts}^\star V_{td}$ is strongly reduced  by the
constraint on the CKM parameters enforced by the measured value of $\eps$.
 In the analysis reported
in this paper,  the same input parameters as in ref.~\cite{rom99}, with the  exception of 
$\Omega_{IB}=0.16\pm 0.03$ which is now taken from~\cite{eck99}, have been used.

The discussion in ref.~\cite{rom99} about the current status of the lattice computation
of the main matrix elements $\langle\pi\pi\vert Q_6\vert K\rangle$ and
$\langle\pi\pi\vert Q_8^{3/2}\vert K\rangle$ can be summarized as follows:
\begin{itemize}
\item At present, the matrix element $\langle\pi\pi\vert Q_6\vert K\rangle$ is not
reliably known from lattice QCD. The results with staggered fermions
are plagued by huge corrections appearing in the operator renormalized
using lattice perturbation theory~\cite{pk99}. Other attempts using Wilson fermions or
domain-wall fermions were unsuccessful so far.

\item The matrix element $\langle\pi\pi\vert Q_8^{3/2}\vert K\rangle$ has been computed
by several groups using different formulations of the lattice fermion action and different
lattice spacing. A substantial agreement of the different determinations was found within
$20\% $ uncertainty. We use the value
\be
\langle\pi\pi\vert Q_8^{3/2}(2\,\mbox{GeV}, \overline{\mbox{MS}}\mbox{-HV})
\vert K\rangle=0.62\pm 0.12\,\mbox{GeV}^3
\ee
for  the operator renormalized at $\mu=2$ GeV in the 't Hooft-Veltman
$\overline{\mbox{MS}}$ scheme. This value corresponds to the $B$ parameter $B_8^{3/2}=0.93\pm
0.18$ for a ``conventional'' quark mass $m_s^{\overline{\mbox{\scriptsize MS}}}+
m_d^{\overline{\mbox{\scriptsize MS}}}=130$ MeV at $\mu=2$ GeV, see eq.~(\ref{eq:bpars}).
\end{itemize}
Using some reasonable assumptions for the less important contributions due to other
operators,  given that the largest uncertainty stems from our ignorance of 
$\langle\pi\pi\vert Q_6\vert K\rangle$, a useful way of presenting the results is given by
\be
\epsp=\left[(-25.3^{+3.1}_{-3.6})\,\mbox{GeV}^{-3}\, \langle\pi\pi\vert Q_6(2\,\mbox{GeV},
\overline{\mbox{MS}}\mbox{-HV})\vert K\rangle - 6.3^{+1.6}_{-1.7}\right]\times 10^{-4}\, ,
\label{eq:lres}
\ee
where the matrix element of $Q_6$ is considered as a free parameter. Notice that the
the two terms in this equation are correlated and should not be varied independently.

\begin{figure}[t]
\vspace{9.0cm}
\includegraphics{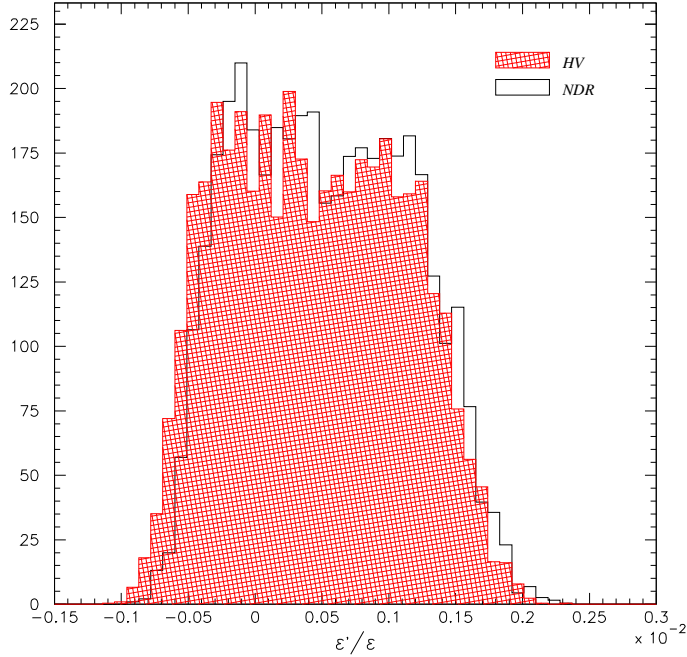}
 \caption{\it
      Distribution of values for $\epp$ using $B_6=1\pm 1$ either in
       $\overline{\mbox{MS}}\mbox{-HV}$ or in $\overline{\mbox{MS}}\mbox{-NDR}$ as
       discussed in the text.
    \label{fig:epsp} }
\end{figure}

In order to compare eq.~(\ref{eq:lres}) with experiments, we have to make some assumption on the value of
$\langle\pi\pi\vert Q_6\vert K\rangle$. We take the central value suggested by the
$VIA$ (or equivalently by the lowest-order $1/N$ expansion),
namely $B_6=1$,
with an uncertainty of $100\%$. This introduce  a renormalization-scheme ambiguity,
since $VIA$ does not allow  a proper definition of the renormalized operators.
For this reason, results obtained by taking two different central values for
$\langle\pi\pi\vert Q_6\vert K\rangle$ (corresponding
either to $B_6^{\mbox{\scriptsize HV}}(2$ GeV)=1 or to
$B_6^{\mbox{\scriptsize NDR}}(2$ GeV)=1 are presented, namely
\be
\begin{tabular}{ll}
$\langle\pi\pi\vert Q_6(2\,\mbox{GeV}, \overline{\mbox{MS}}\mbox{-HV})
\vert K\rangle=-0.4\pm 0.4\,\mbox{GeV}^3$ &
($B_6^{\mbox{\scriptsize HV}}=1\pm 1$), \\
$\langle\pi\pi\vert Q_6(2\,\mbox{GeV}, \overline{\mbox{MS}}\mbox{-HV})
\vert K\rangle=-0.6\pm 0.6\,\mbox{GeV}^3$ &
($B_6^{\mbox{\scriptsize NDR}}=1\pm 1$).
\label{example}
\end{tabular}
\ee

In the two cases, we obtain
\be
\begin{tabular}{ll}
$\epp = (4.6^{+7.7}_{-7.2}\pm 0.4)\times 10^{-4}$ & Monte Carlo\\
$-13\times 10^{-4} \le \epp \le 26\times 10^{-4}$ & scanning\\
\\
$\epp = (8.1^{+10.3}_{-9.5}\pm 0.3)\times 10^{-4}$ & Monte Carlo\\
$-13\times 10^{-4} \le \epp \le 37\times 10^{-4}$ & scanning\\
\label{eq:res}
\end{tabular}
\ee
The difference of the two results, contrary to what is often stated in the literature,
is not the uncertainty associated to the renormalization-scheme dependence, but to a
different choice of the value of the matrix element in a given  scheme  (the
HV scheme in the example of eq.~(\ref{example})).
At the NLO, the scheme dependence comes from higher-order
corrections only and its effect is estimated by the second error given in eq.~(\ref{eq:res}).
The two figures  of eq.~(\ref{eq:res}) correspond really to two different choices of
the unknown value of $\langle\pi\pi\vert Q_6\vert K\rangle$ at a given scale ($\mu=2$ GeV)
and in a well-defined scheme ($\overline{\mbox{MS}}\mbox{-HV}$).
On the contrary, the two distributions of values for  $\epp$  in fig.~\ref{fig:epsp}
include, for the same choice of  $\langle\pi\pi\vert Q_6\vert K\rangle$,
 two different ways to match Wilson coefficients and matrix elements
for estimating the real  scheme dependence due to higher-order terms. Both distributions refer to the case
$B_6(2\,\mbox{GeV},\overline{\mbox{MS}}\mbox{-HV})=1\pm 1$. The large error on
the matrix elements of $Q_6$ obviously dominates the final uncertainty on $\epp$ and flattens
these distributions. In ref.~\cite{mun99} a more optimistic error for the $B$ parameter
($B_6(2\,\mbox{GeV},\overline{\mbox{MS}}\mbox{-HV})=1\pm 0.2$) was assumed.

\section{Comparison with data}
Many of the Standard Model predictions shown in fig.~\ref{fig:comp} are
below the present experimental world average. What does this imply? There are three
legitimate answers:
\begin{enumerate}
\item {\em There is nothing wrong!}  For some specific choice of the input parameters,
all the different approaches are able to 
reproduce to some extent the experimental data.  In some cases the agreement seems to arise naturally
from the calculation~\cite{nar00,xbos}.  In other cases,  this requires the adjustment 
of a few parameters~\cite{tri00} or a wise choice of several of them
(often at  the edge of the allowed range of values)~\cite{mun99,rom99}.
It is puzzling that most  of the approaches, which suffer from   intrinsic and irreducible 
uncertainties coming from the  model dependence 
of the results,  are in good agreement with the data.
In the case of refs.~\cite{rom99,mun99}, instead,
this requires that all the quantities on  which we have a poor control conspire in the 
direction to increase the theoretical value of $\epp$.
Thus,   although unlikely in our opinion,  the possibility that there 
is nothing wrong is not excluded.
It may also well be that some of the models
are indeed able to describe the underlying strong dynamics.

\item {\em There is something  missing in the  computation of the matrix elements.}
The long-standing problem of explaining the
$\Delta I=1/2$ rule suggests that some important  dynamical effect is at work  in $K\to
\pi\pi$ $I=0$ decays.  Unfortunately, contrary to some old claim, there  is no simple relation between the
$CP$-conserving and $CP$-violating decays, which could explain
the large value of $\epp$  on the basis of the enhancement 
of the $\Delta I=1/2$  amplitude. Indeed, it would be very interesting
if a common dynamical mechanism could explain both of them. In terms of Wick contractions
in  the matrix elements, such a mechanism could be possibly provided by a large contribution
from eye diagrams (aka penguin contractions)~\cite{ciu99}. From the lattice estimates, by taking $B_6$ as a free
parameter, we can reproduce the experimental $\epp$ with
$B_6^{\mbox{\scriptsize HV}}(2\,\mbox{GeV})\sim 2.4$.
As we have seen, all non-perturbative methods are affected by  theoretical and/or
computational problems which limit their accuracy. 
Among them,  the models based on the chiral expansion also support the existence of some
correlation between the $\Delta I=1/2$ rule and $\epp$ which is at least in qualitative
agreement with the observations. A possible exception is that of ref.~\cite{bel99}. Thus we conclude
that  a real quantitative explanation is still to come. 

\item {\em Hadronic matrix elements are fine. New physics is at work.} 
If the theoretical calculations which gives low values
for $\epp$ are correct, there is room for new physics effects. It is not difficult
to imagine new sources of $CP$ violation. In supersymmetry, for example, there are even
too many. The problem is that we must find  a model for new physics  such as to
obtain a sizeable contribution to $\epp$ while remaining within 
the stringent constraints imposed by $\varepsilon$ and by
 other measured quantities.
This problem can be circumvented, so that for instance supersymmetry is potentially able (with some
special assumptions) to
produce the required effect on $\epp$ still fulfilling the other phenomenological
constraints~\cite{ant}.
\end{enumerate}

At present, our preferred answer is the second one. Hopefully, improvements in
non-perturbative techniques and a further insight in kaon phenomenology will
clarify the mechanism responsible for the ``large'' value of $\epp$ and
its connections with the $\Delta I=1/2$ rule.

\section*{Acknowledgments}
It is a pleasure to acknowledge valuable discussions with A.~Buras,
E.~Franco, G.~Isidori and V.~Lubicz on the subject of this work.
M.C. acknowledges a partial support by the German Bundesministerium
f\"ur Bildung und Forschung under contract 06TM874 and 05HT9WOA0.


\begin{thebibliography}{99}
\bibitem{KTeV} A.~Alavi-Harati {\it et al.}, Phys. Rev. Lett. {\bf 83} (1999) 22.
\bibitem{NA48} A.~Ceccucci, CERN Particle Physics Seminar (29 February 2000),
 http://www.cern.ch/NA48/Welcome/images/talks/cern00/talk.ps.gz.
\bibitem{dagostini} G.~D'Agostini,  CERN-EP-99-139,  hep-ex/9910036. 
\bibitem{d0d2} E.~Chell and M.G.~Olsson, Phys. Rev. {\bf D48} (1993) 4076.
\bibitem{bg87} A.J.~Buras and J.-M.~G\'erard, Phys. Lett. {\bf B192} (1987) 156.
\bibitem{ds1} A.J.~Buras, M.~Jamin and M.E.~Lautenbacher and P.H.~Weisz,
 Nucl. Phys. {\bf B400} (1993) 37;
 A.J.~Buras, M.~Jamin and M.E.~Lautenbacher, {\it ibid.} 75;
 M.~Ciuchini, E.~Franco, G.~Martinelli and L.~Reina,
 Nucl. Phys. {\bf B415} (1994) 403.
\bibitem{kns99} Y.-Y.~Keum, U.~Nierste and A.I.~Sanda, Phys. Lett. {\bf B457}
 (1999) 157.
\bibitem{giusti} A.~Donini {\it et al.}, Phys. Lett. {\bf B470} (1999) 233.
\bibitem{k99} M.~Ciuchini {\it et al.}, hep-ph/9910237, to appear in: Proc.
of KAON `99, June 21--26 1999, Chicago, USA.
\bibitem{rom99} M.~Ciuchini {\it et al.}, Nucl. Phys {\bf B573} (2000) 201.
\bibitem{luscher} M.~L\"uscher, Les Houches Lectures on ``Advanced Lattice QCD",
 hep-lat/9802029, and refs. therein.
\bibitem{wittig} J.~Garden {\it et al.},  Nucl.~Phys. B571 (2000) 237.
\bibitem{mt} B.Yu.~Blok and M.A.~Shifman, Sov. J. Nucl. Phys. {\bf 45} (1987) 522;
L.~Maiani and M.~Testa, Phys. Lett. {\bf B245} (1990) 585.
\bibitem{ppp} E.~Pallante and A.~Pich, Phys. Rev. Lett. {\bf 84} (2000) 2568;
 E.~Paschos, hep-ph/9912230.
\bibitem{rom95} M.~Ciuchini, E.~Franco, G.~Martinelli and L.~Silvestrini,
Phys. Lett. {\bf B380} (1996) 353
\bibitem{ll} L.~Lellouch and M.~L\"uscher, hep-lat/0003023.
\bibitem{sacring} Talk by C.T.~Sachrajda at the Ringberg Workshop on Non-perturbative
Methods, Ringberg, April 2-8, 2000; G.~Martinelli and C.T.~Sachrajda, to appear.
\bibitem{mun99} S.~Bosch {\it et al.}, Nucl. Phys. {\bf B565} (2000) 3.
\bibitem{mun93} A.J.~Buras, M.~Jamin and M.E.~Lautenbacher, Nucl. Phys. {\bf B408}
(1993) 209.
\bibitem{dor99} T.~Hambye, G.O. K\"ohler, E.A.~Paschos and P.H.~Soldan,
 Nucl. Phys. {\bf B564} (2000) 391.
\bibitem{bbg} W.A.~Bardeen, A.J.~Buras and J.-M.~G\'erard, Phys. Lett.
 {\bf B180} (1986) 133;  W.A.~Bardeen, A.J.~Buras and J.-M.~G\'erard,
 Phys. Lett. {\bf B192} (1987) 138; W.A.~Bardeen, A.J.~Buras and J.-M.~G\'erard,
 Phys. Lett. {\bf B211} (1988) 343.
\bibitem{tri00} S.~Bertolini, J.O. Eeg and M.~Fabbrichesi, hep-ph/0002234
 and refs. therein..
\bibitem{bel99} A.A.~Bel'kov, G.~Bohm, A.V.~Lanyov and A.A.~Moshkin, hep-ph/9907335.
\bibitem{che99} H.-Y.~Cheng, hep-ph/9911202.
\bibitem{genfac} M.~Neubert and B.~Stech,  CERN-TH-97-099,
in Heavy flavours II,
A.~Buras and  M.~Lindner eds., p. 294, hep-ph/9705292.
\bibitem{har99} M.~Harada {\it et al.}, hep-ph/9910201.
\bibitem{gavela} M.B.~Gavela {\it et al.}, Phys.~Lett. B211 (1988) 139. 
\bibitem{eck99} G.~Ecker, G.~M\"uller, H.~Neufeld and A.~Pich, Phys. Lett.
 {\bf B477} (2000) 88.
\bibitem{gv99} S.~Gardner and G.~Valencia, Phys. Lett. {\bf B466} (1999) 355.
\bibitem{mun00} A.J.~Buras {\it et al.}, Phys. Lett. {\bf B480} (2000) 80.
\bibitem{nar00} S.~Narison, hep-ph/0004247.
\bibitem{xbos} J.~Bijnens and J.~Prades, JHEP {\bf 0001} (2000) 002;
J.~Bijnens and J.~Prades, hep-ph/0005189.
\bibitem{ckm} P. Paganini, F. Parodi, P. Roudeau and A. Stocchi,
 Phys. Scripta {\bf V. 58} (1998) 556;
 F. Parodi, P. Roudeau and A. Stocchi, Nuovo Cim. {\bf 112A} (1999) 833;
 F. Caravaglios, F. Parodi, P. Roudeau and A. Stocchi, hep-ph/0002171;
 S. Mele, Phys. Rev. {\bf D59} (1999) 113011;
 A. Ali and D. London, Eur. Phys J. {\bf C9} (1999) 687;
 M.~Bargiotti {\it et al.}, La Rivista del Nuovo Cimento, Vol. 23, N.3 (2000), 1;
 S. Plaszczynski, M.-H. Schune, hep-ph/9911280.
\bibitem{pk99}  S.~Sharpe {\em et al.},
Phys. Lett. {\bf 192B} (1987) 149;  S.~Sharpe and A.~Patel, Nucl. Phys. {\bf B417}
(1994) 307;
N.Ishizuka and Y.~Shizawa, Phys. Rev. {\bf D49} (1994) 3519;
D.~Pekurovsky and G.~Kilcup, Nucl. Phys. {\bf Proc. Suppl. 63} (1998) 293;
D.~Pekurovsky and G.~Kilcup, hep-lat/9812019.
\bibitem{ciu99} M.~Ciuchini, E.~Franco, G.~Martinelli and L.~Silvestrini,
 hep-ph/9909530, to appear in: Proc. of KAON `99, June 21--26 1999, Chicago, USA.
\bibitem{ant} A.~Masiero and O.~Vives, hep-ph/0001298 and refs.
therein.
\end{thebibliography}
\end{document}